\documentclass[preprint,12pt,3p]{elsarticle}
\usepackage{graphicx}
\usepackage{amssymb}
\usepackage{lineno}
\usepackage{caption}
\usepackage{subcaption}
\usepackage{multirow}
\usepackage{caption}
\usepackage{subcaption}
\usepackage{amsmath}
\usepackage[T1]{fontenc}

\usepackage[hidelinks]{hyperref}

\journal{WSTLU2021}

\begin{document}

\begin{frontmatter}

\title{Do e-scooters fill mobility gaps and promote equity before and during COVID-19? A spatiotemporal analysis using open big data}



\author[UF1]{Xiang Yan\corref{cor1}}
\author[UF2]{Wencui Yang}
\author[UF1]{Xiaojian Zhang}
\author[UF1]{Yiming Xu}
\author[UF2]{Ilir Bejleri}
\author[UF1]{Xilei Zhao\corref{cor1}}
\cortext[cor1]{Corresponding author. Address: 1949 Stadium Rd, Gainesville, FL 32611. Email: xiangyan@ufl.edu.}

\address[UF1]{Department of Civil and Coastal Engineering, University of Florida, Gainesville, FL}
\address[UF2]{Department of Urban and Regional Planning, University of Florida, Gainesville, FL}

\begin{abstract}
The growing popularity of e-scooters and their rapid expansion across urban streets has attracted widespread attention. A major policy question is whether e-scooters substitute existing mobility options or fill the service gaps left by them. This study addresses this question by analyzing the spatiotemporal patterns of e-scooter service availability and use in Washington DC, focusing on their spatial relationships with public transit and bikesharing. Results from an analysis of three open big datasets suggest that e-scooters have both competing and complementary effects on transit and bikesharing services. The supply of e-scooters significantly overlaps with the service areas of transit and bikesharing, and we classify a majority of e-scooter trips as substitutes to transit and bikesharing uses. A travel-time-based analysis further reveals that when choosing e-scooters over transit, travelers pay a price premium and save some travel time. The price premium is greater during the COVID-19 pandemic but the associated travel-time savings are smaller. This implies that public health considerations rather than time-cost tradeoffs are the main driver for many to choose e-scooters over transit during COVID. In addition, we find that e-scooters complement bikesharing and transit by providing services to underserved neighborhoods. A sizeable proportion (about 10 percent) of e-scooter trips are taken to connect with the rail services. Future research may combine the big-data-based analysis presented here with traditional methods to further shed light on the interactions between e-scooter services, bikesharing, and public transit.
\end{abstract}

\end{frontmatter}

\section{Introduction}
\label{S:1}

As a new form of micromobility, e-scooters (i.e., shared electric scooters) quickly emerged as a popular travel mode among urban residents. In 2018, just the second year after e-scooters first appeared on urban streets, the number of e-scooter trips (38.5 million) surpassed that of station-based bikeshare trips (36.5 million) in the U.S. \citep{NACTO2019}. Many people have predicted the e-scooter market to further grow, as evident by the billions of venture capital being poured into major e-scooter startups such as Bird, Lime, and Spin. This prediction is supported by travel-behavior findings from the 2017 U.S. National Household Travel Survey data, which showed that about 46\% (or even a higher percentage due to under-reporting of short trips) of personal trips are three miles or less \citep{NHTS2017}. E-scooters, if priced properly and supported by both government policies and adequate infrastructure, have a great potential to be used for these trips. Currently, a majority (more than 70\%) of trips of three miles or below were driven, and less than two percent of them were made by bicycles \citep{NHTS2017}.

The popularity of e-scooters and their rapid expansion across urban streets lead to mixed reactions. Some acclaim the invention of a new form of urban mobility, seeing the petrol-free e-scooters as an enabler of sustainable travel and lifestyles. Supporters of e-scooters believe that they can reduce car use and traffic congestion, facilitate transit access by expanding the catchment area of transit stops, and promote accessibility for people living in communities with fewer travel options \citep{clewlow2019micro}. However, critics raise a variety of concerns including improper parking \citep{brown2020impeding}, travel safety, vandalism, and the lack of infrastructure to accommodate e-scooter use. In addition, it is possible that e-scooters mainly replace walking, station-based bikeshare, and transit trips, thus undermining other sustainable modes of travel. These concerns suggest the need for policy guidelines and measures to ensure that micromobility can better fit into a city's transportation ecosystem.

As cities around the world deliberate how to manage and regulate e-scooters, there is an urgent need to learn how e-scooters interact with existing mobility options especially public transit and station-based bikesharing (to be referred to as bikesharing for simplicity). On the one hand, e-scooters are close substitutes to these options. A big worry to transportation professionals in the U.S. is that e-scooters may attract away transit riders, further exacerbating the trend of declining transit ridership \citep{wasserman2020s}. On the other hand, e-scooters may be complementary to public transit by facilitating first-mile/last-mile connections to transit stops \citep{oeschger2020micromobility}. In addition, affordable e-scooters services may enhance transportation for neighborhoods that receive inadequate transit services and lack convenient access to bikesharing systems. Until now, the question of whether e-scooters replace existing sustainable travel modes or fill the mobility gaps left out by them is largely unexplored. 

To fill this knowledge gap, this paper presents a spatiotemporal analysis of the interaction between e-scooters services and two existing travel modes (i.e., public transit and bikesharing) in the city of Washington DC. We address the following research questions:

\textbf{RQ1.} How much does the supply of e-scooter services overlap with that of bikesharing and transit services? 

\textbf{RQ2.} Do e-scooter trips substitute or complement transit and bikesharing trips? If both effects exist, what proportion of e-scooters fall into each category? 

\textbf{RQ3.} For e-scooter trips that potentially substitute transit use, how do their travel times compare with those of the transit alternative?

Answers to \textbf{RQ1} and \textbf{RQ2} shed light on the competing and complementary relationships between e-scooters and existing modes from both the supply and demand perspectives. A further examination of travel-time differences (\textbf{RQ3}) provides a more nuanced understanding of the degree of competition. For instance, if an e-scooter trip costs much less time than the fastest transit alternative, argument can go both ways for whether it complements or competes with public transit. Our study context is the city of Washington DC, which has both robust transit and bikesharing systems and a mature e-scooter market. Since the recent COVID-19 pandemic has greatly disrupted how people travel and caused significant disruptions to public transit, we conduct the analysis for two periods of time: before and during COVID-19. The results can not only inform policymaking on how to better integrate e-scooters into the existing transportation system but also shed light on how e-scooters may enhance transportation resilience in a time of crisis. 


\section{Literature Review}
\subsection{E-scooter Users and Trip Characteristics}
E-scooters were first launched in the city of Santa Monica, California in late 2017 and were quickly expanded to other cities such as San Francisco and Washington DC in early 2018 \citep{clewlow2019micro}. By 2020, hundreds of cities worldwide have e-scooters on their streets. Some U.S. cities conducted surveys to collect the demographic and socioeconomic information of the e-scooter users \citep{Alington2019,Portland2018,SF2019}. The survey results generally reported that a higher proportion of e-scooter users were male; and compared to the general population, they tend to be younger, college educated, and have higher income. Also, minority populations including Blacks, Hispanics, and Asians tend to be underrepresented, although this may result from sampling bias. These results mirror findings from studies on other new mobility options such as bikesharing and ridesourcing \citep{buck2013bikeshareuser,young2019uberuser}.

Analysts have found that, on average, e-scooter trips are quite short. In most cities, the average trip distance was below two miles and the average trip duration was below 20 min \citep{bai2020dockless,zou2020exploratory,SF2019,Alington2019}. The average trip distance and duration were even lower in cities (e.g., Washington DC, San Francisco, and Austin) where e-scooters were disproportionately placed at tourist attractions, possibly due to the use of e-scooters for leisure trips. Nonetheless, a majority of e-scooters trips were made for utilitarian purposes such as commuting, shopping or errands, and to connect with public transit \citep{Alington2019,Portland2018,SF2019}. Several studies suggested that e-scooters were more heavily used in the afternoon and early evening, although a slight rise in usage was often observed in the morning peak hours \citep{mckenzie2019spatiotemporal,zou2020exploratory,Indianapolis2019}. Also, weekends saw increased e-scooter trip-making over weekdays. Finally, e-scooter trip origins and destinations were usually concentrated, clustering around downtown areas, university campuses, and tourist attractions \citep{bai2020dockless,zou2020exploratory,SF2019}.

\subsection{Spatiotemporal Comparison of E-Scooter and Bikesharing Trips}
Up until 2018 before e-scooters gained widespread popularity, bikesharing (mainly station-based bike share) trips have grown steadily in recent years \citep{NACTO2019}. Meanwhile, while dockless bikes proliferated around the world, especially in Asia \citep{xu2019unravel,SONG2021101566}, they experienced limited growth in the U.S. Given this context, the discussion here focuses on comparing e-scooter sharing with station-based bikesharing only. E-scooter sharing differs from bikesharing in two notable ways. First, unlike bikesharing that requires users to pick up and drop off vehicles at fixed stations, e-scooters can be placed at anywhere and so offer a more flexible and convenient service. Second, bikesharing programs are mostly government funded in North America, whereas e-scooters are completely operated by private firms so far. While e-scooters and bikeshare are expected to attract travelers with similar demographics, survey results suggested that a very small percentage (usually less than 5\%) of e-scooter trips replaced bikeshare trips \citep{Portland2018,SF2019,Alington2019}.

To understand the differences between e-scooter sharing and bikesharing services, several studies have compared their spatial and temporal usage patterns \citep{mckenzie2019spatiotemporal,younes2020comparing,zhu2020understanding}. The McKenzie study revealed some degree of similarities but significant differences between e-scooter trips and bikesharing trips both in the temporal and spatial dimensions. Notably, bikesharing trips taken by nonmembers and e-scooter trips had a similar spatial distribution, but they differed substantially in their temporal patterns. Bikesharing usage by members differed from scooter usage significantly both spatially and temporally. These results suggested that e-scooters replaced some bikesharing trips but the two types of services were often used for different purposes \citep{mckenzie2019spatiotemporal}. Also analyzing the Washington DC data, \citet{younes2020comparing} confirmed some of the McKenzie study findings; they further showed that weather is less of a disutility for e-scooter users than bikeshare users. \citet{zhu2020understanding} found that, in Singapore, e-scooter trips was more spatially concentrated than bikeshare trips. They also suggested that e-scooters had a higher utilization rate but faced a rebalancing challenge due to the need for battery charging. 

\subsection{The Relationship Between E-Scooters and Transit}
Like other new forms of shared mobility such as ridesourcing and bikesharing, e-scooters can be both an enabler and disruptor to public transit. A major deterrent to transit use is the ``first-mile/last-mile problem,'' which refers to the inability of some travelers to reach transit stops that are too far way \citep{boarnet2017first}. By providing a faster and convenient alternative to walking, e-scooters can address this problem and hence augment transit use. Evaluations of e-scooter pilot programs showed that many travelers indeed used e-scooters to connect to transit. For instance, 34\% of survey respondents in San Francisco and 18\% of e-scooter riders in Arlington, Virginia, respectively reported using e-scooters to get to or from public transit \citep{SF2019}. Moreover, e-scooters can be deployed in large volumes and widely distributed across space, hence providing access to places that are unserved or underserved by transit. To achieve this goal, some cities have required or incentivized e-scooter companies to place a required amount of vehicles in designated equity zones. On the other hand, e-scooters may replace some transit trips \citep{NACTOGuide2019}. Survey results suggested that had e-scooters not been available for the last scooter trip, between 5\%--11\% of respondents would have taken public transit \citep{Alington2019,SF2019,Portland2018}. Also, after the adoption of e-scooter services, individuals generally reported using transit less frequently than before \citep{Portland2018,Omaha2020}. 


Overall, these pilot programs suggest that e-scooters have both a complementary and substitutionary effect on public transit. This finding is consistent with results of previous studies that examined the impact of bikesharing systems on transit \citep{martens2004bicycle,martin2014evaluating,ma2015bicycle,campbell2017sharing}. Bikesharing studies produced two additional insights that are likely to be applicable for e-scooter services. First, travelers are much more likely to integrate a bikesharing trip with rail than with bus \citep{martens2004bicycle}. This is likely a main explanation for studies to find that bikesharing has a complementary effect on metro rail but a substitutionary effect on bus services \citep{ma2015bicycle,campbell2017sharing}. Second, how exactly the complementary and substitutionary effects of bikesharing on public transit play out largely depends on the urban context. Martin and Shaheen showed that bikesharing serves prominently as a first-mile/last-mile facilitator in areas with less intensive transit network but replaces many short transit trips in high-density, transit-rich areas \citep{martin2014evaluating}.

\section{Data}

Washington DC is one of the pioneer cities that brought shared e-scooters onto the streets. The District Department of Transportation has established the terms and conditions for e-scooter operations, and it requires each vendor to provide an Application Programming Interface (API) to share data with the public \citep{mckenzie2020urban}. The APIs provide data in the General Bikeshare Feed Specification (GBFS) format, an open data standard for bikeshare system availability. The GBFS data attributes include vehicle (bike or e-scooter) ID, latitude and longitude of vehicle location, whether the vehicles is reserved or disabled, battery level, etc. In addition to the GBFS data, two other open data sources used here include the General Transit Feed Specification (GTFS) data published by the Washington Metropolitan Area Transit Authority and the real-time system data published by Capital Bikeshare. The GTFS data provide information on transit routes, stops, and schedules. The real-time bikesharing system data contain the geographic information of bikesharing stations and the number of shared bikes available at each station at a given time point. 

We developed a Python program to scrap the GBFS data at a one-minute time interval (vendors update their APIs at a different time intervals, ranging from one minute to 10 min). The raw GBFS data scrapped from the APIs indicate the supply of e-scooters in the city at a given time point; and by examining how the ``bike ID'' field changes over time, one may also infer trips from the GBFS data. Figure 1 illustrates the e-scooter availablility and e-scooter trip data used in this study. Since the types of ``bike ID'' reported by each e-scooter vendor can differ, the trip information to be extracted differs across vendors \citep{xu2020micromobility}. For vendors (e.g., Jump, Skip, Spin) that assign a consistent ID for the same scooter over time, one can infer trip origin-destination pairs; for vendors (Bird, Lime, Lyft, Razor) that assign a dynamic ID for the same e-scooter, one can only unlinked trip origins and destinations. Xu et al. (2020) provides a detailed description of the trip inference algorithms adopted in this study. Given the nature of the data, we used data from all vendors when examining the supply of e-scooters (\textbf{RQ1}) and data from vendors that assign consistent IDs to scooters when examining e-scooter trips (\textbf{RQ2} and \textbf{RQ3}). Omitting trips served by several vendors from the analysis is not likely to bias the results, as we expect e-scooters services operated by different vendors to have similar substituting or complementary effects on transit or bikesharing.

Note that some trips may be falsely identified due to GPS error or vehicle recharging. To address this issue, we have excluded trips that are shorter than 0.02 mile or longer than ten miles, are shorter than five min or longer than 90 min, or have an average travel speed above 20 miles per hour \citep{zou2020exploratory}. To compare the differences of trip patterns before and during COVID-19, we pick a typical week from each period for the subsequent analysis. Specifically, we analyze the week of July 15-21, 2019 for the pre-COVID period and the week of June 15-21, 2020 for the COVID-19 period. A comparison of the results for the before- and during-COVID weeks suggests that travelers have taken less e-scooter trips during COVID-19, largely as a result of reduced travel. On the other hand, e-scooter trips taken during COVID-19, on average, are longer in duration and distance (see Table 2 below). This indicates that people who use transit for longer trips before may have switched to e-scooters, as the estimated transit ridership in Washington DC declined for over 60 percent in the during-COVID week.\footnote{See estimates by the Transit app at https://transitapp.com/coronavirus.} The next section presents a detailed analysis of the e-scooter availability and trip data to further shed light on e-scooter's relationship with transit and bikesharing systems.

\begin{figure}
    \centering
    \begin{subfigure}[h]{0.48\textwidth}
      \centering
      \includegraphics[width=\textwidth]{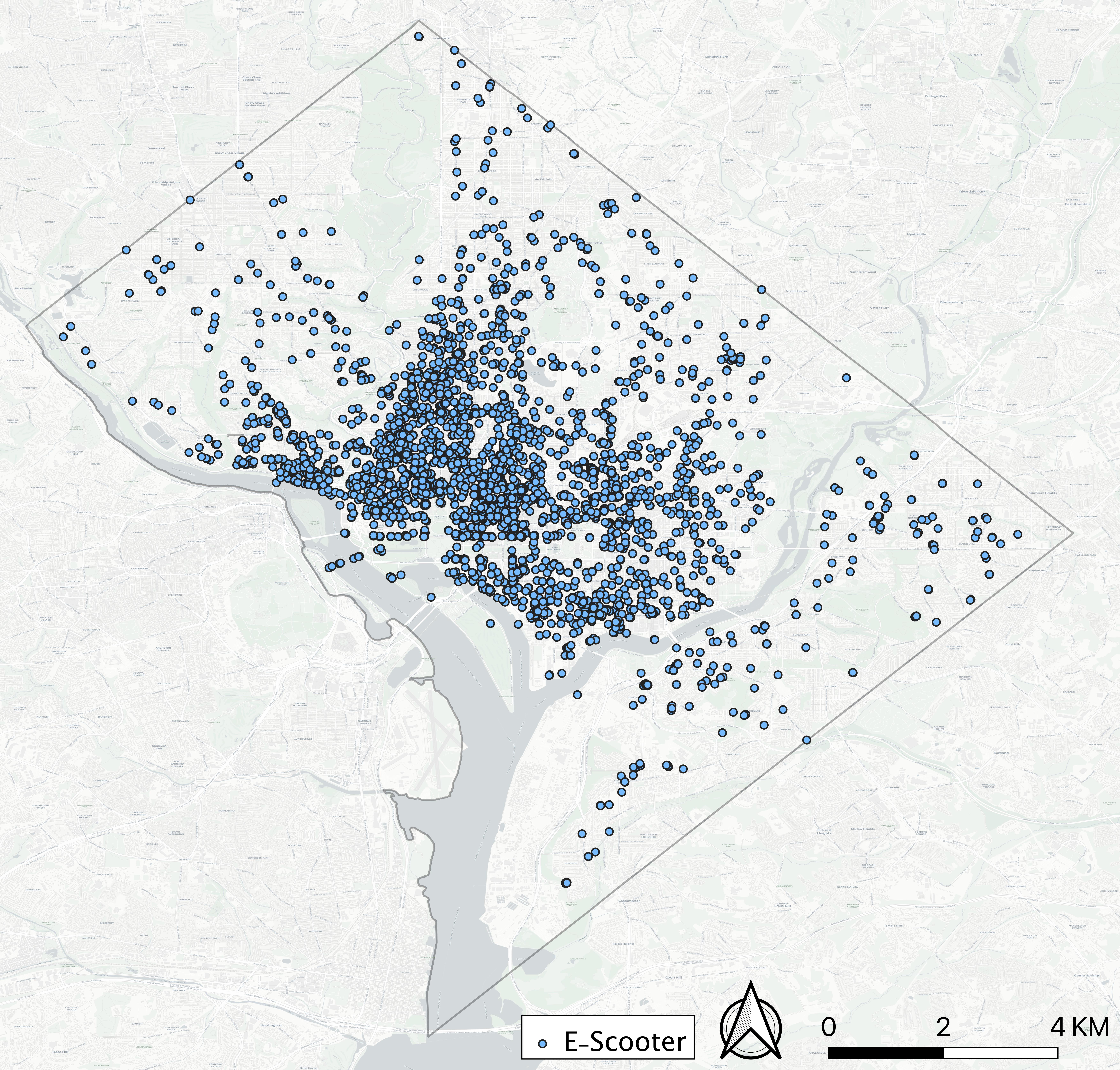}
      \caption{Available e-scooters at 7:00 am on 6/19/2020}
    \end{subfigure}
    \hfill
    \begin{subfigure}[h]{0.48\textwidth}
      \centering
      \includegraphics[width=\textwidth]{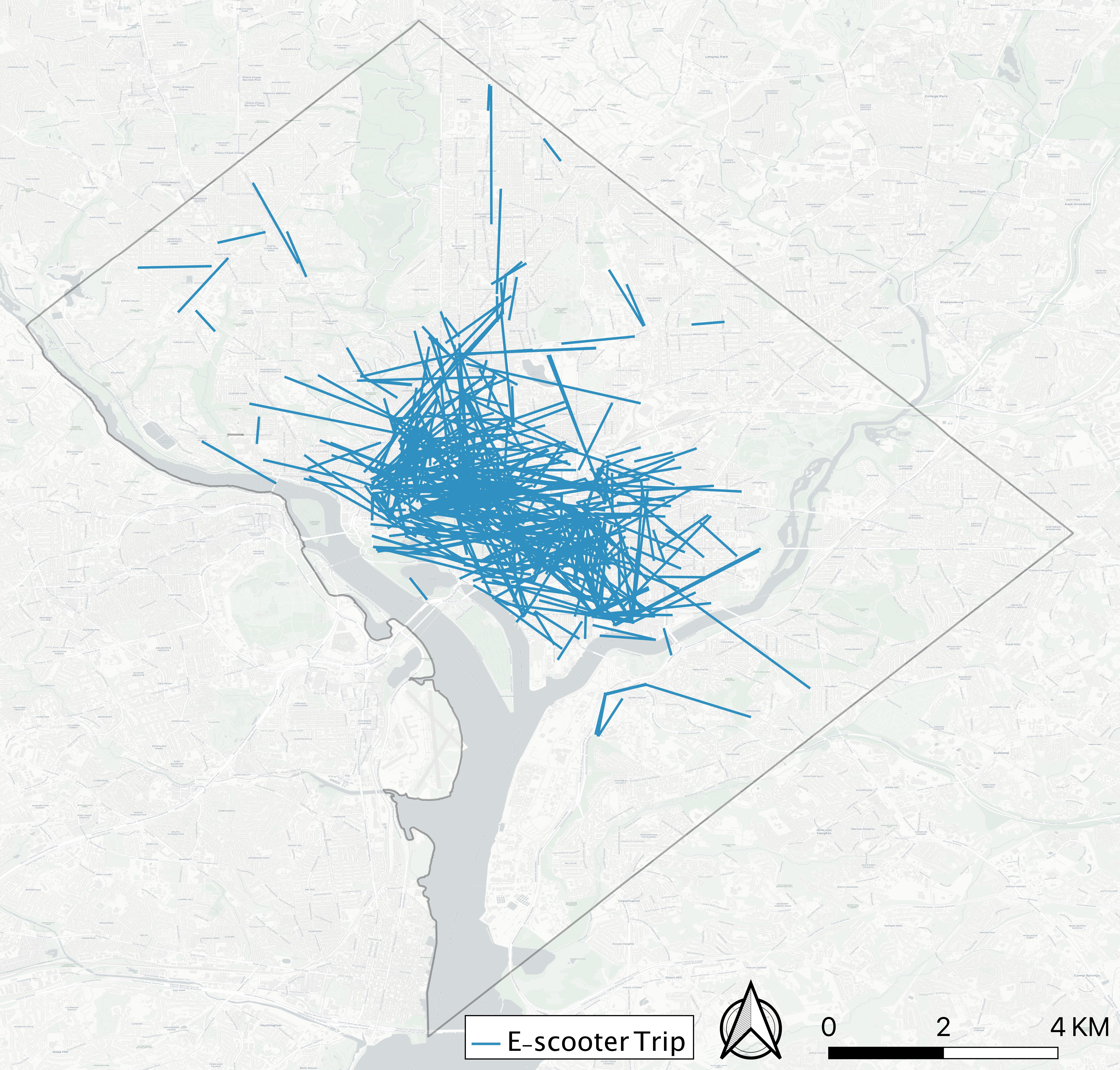}
      \caption{E-scooter trips happening from 7 to 8 am on 6/19/2020}
    \end{subfigure}
    
     \caption{An illustration of the GBFS applied in this study}
        \label{fig:1}
\end{figure}

\section{Analysis and Results}
\subsection{Supply of E-Scooters Versus Bikesharing and Transit}
The analysis of the supply side is fundamental to understand whether e-scooters compete or complement bikesharing and public transit. If these services are offered at the same geographic locations, they would be competing for the same customer base; if e-scooters extend mobility services to neighborhoods with low access to the bikesharing and transit systems, e-scooters would be complementing existing travel modes. However, the relationship in question eludes a dichotomous classification, as both scenarios are likely to exist. To generate more nuanced knowledge on the e-scooters' relationship with bikesharing and public transit, a meaningful path of inquiry is to examine the intensity of these mobility services available at different locations (\textbf{RQ1}). Since the service supply varies throughout the day, one shall also consider the temporal variations. In this study, we compare the supply of the three mobility options at four different points in time: 7:00 am (morning peak hour), 12:00 pm (midday), 5:00 pm (afternoon peak), and 8:00 pm (early evening). We measure the supply of e-scooters and bikesharing across space with the available vehicles at different locations. We measure the supply of transit services at each transit stop by counting the number of vehicles passing by in the following hour (e.g., 7:00 am to 8:00 am). 



We use kernel density to measure the intensity of mobility-service supply across the city, a commonly applied approach to measure accessibility to spatially distributed attractions or resources such as hospitals and parks \citep{wang2012measurement,zhang2011modeling}. The ``resources'' considered here are transit stops, bikesharing stations, and e-scooters. The kernel density approach assumes that the level of accessibility to a given feature (e.g., e-scooter) decreases as the distance to it increases, and the value of accessibility reaches zero at a presumed threshold distance. This threshold distance is usually specified as the service radius of the feature being examined. Here we set this value as a quarter mile for transit stops, one eighth of a mile for bikesharing stations, and one sixteenth of a mile for e-scooters. These values are assumed to be the maximum walking distances for people to ride public transit, use a shared bike, and to find an e-scooter. In addition, a population field can be specified to weight some features more heavily than others. We set the population field as the number of vehicles (a rail train is counted as five vehicles) passing by in an hour for transit stops, the number of available bikes for bikesharing stations, and one for e-scooters. 


\begin{figure}[!ht]
     \centering
     \begin{subfigure}[h]{0.32\textwidth}
         \centering
         \includegraphics[width=\textwidth]{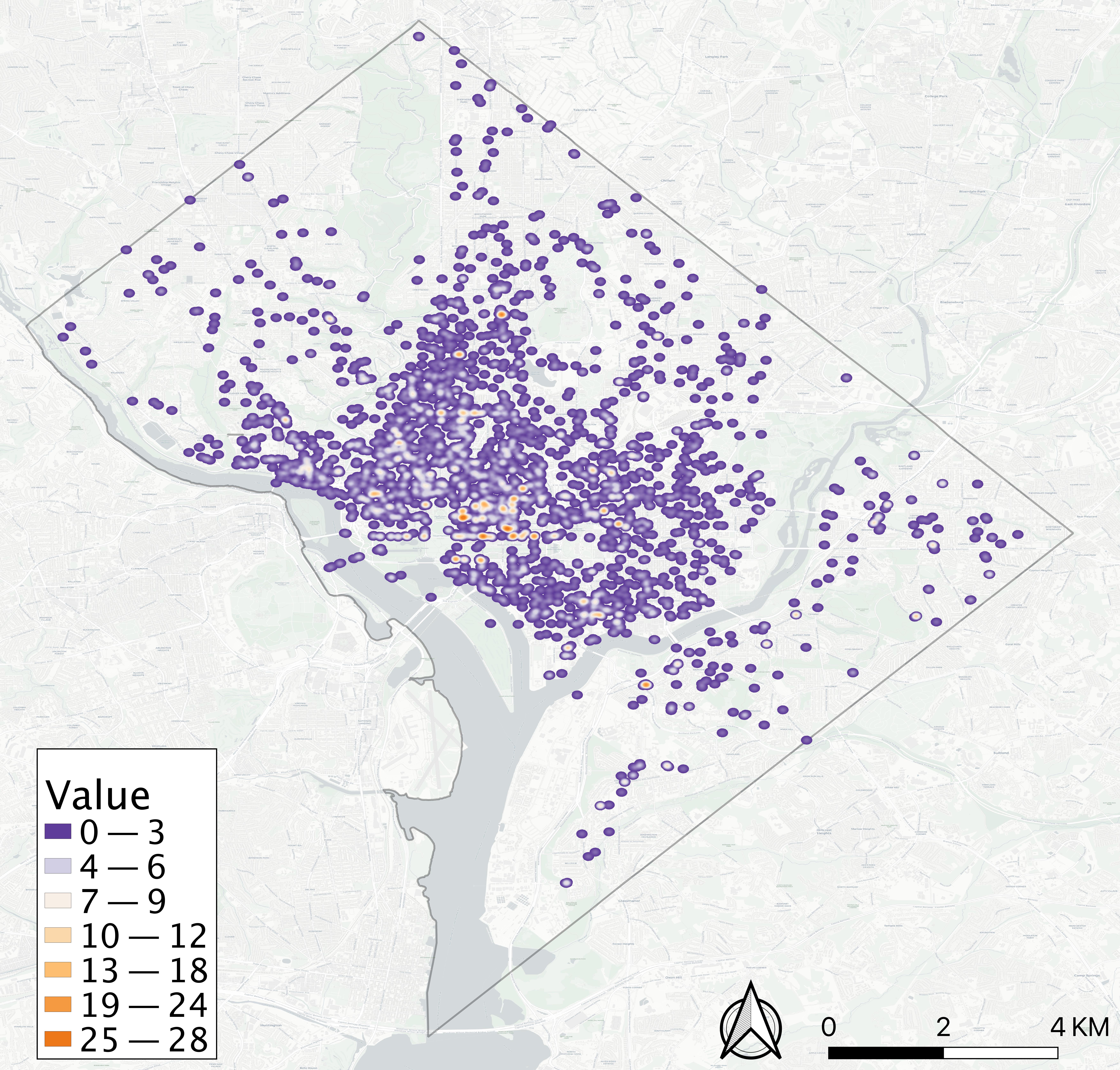}
         \caption{E-scooter, before COVID}
         \label{fig:1a}
     \end{subfigure}
     \hfill
          \begin{subfigure}[h]{0.32\textwidth}
         \centering
         \includegraphics[width=\textwidth]{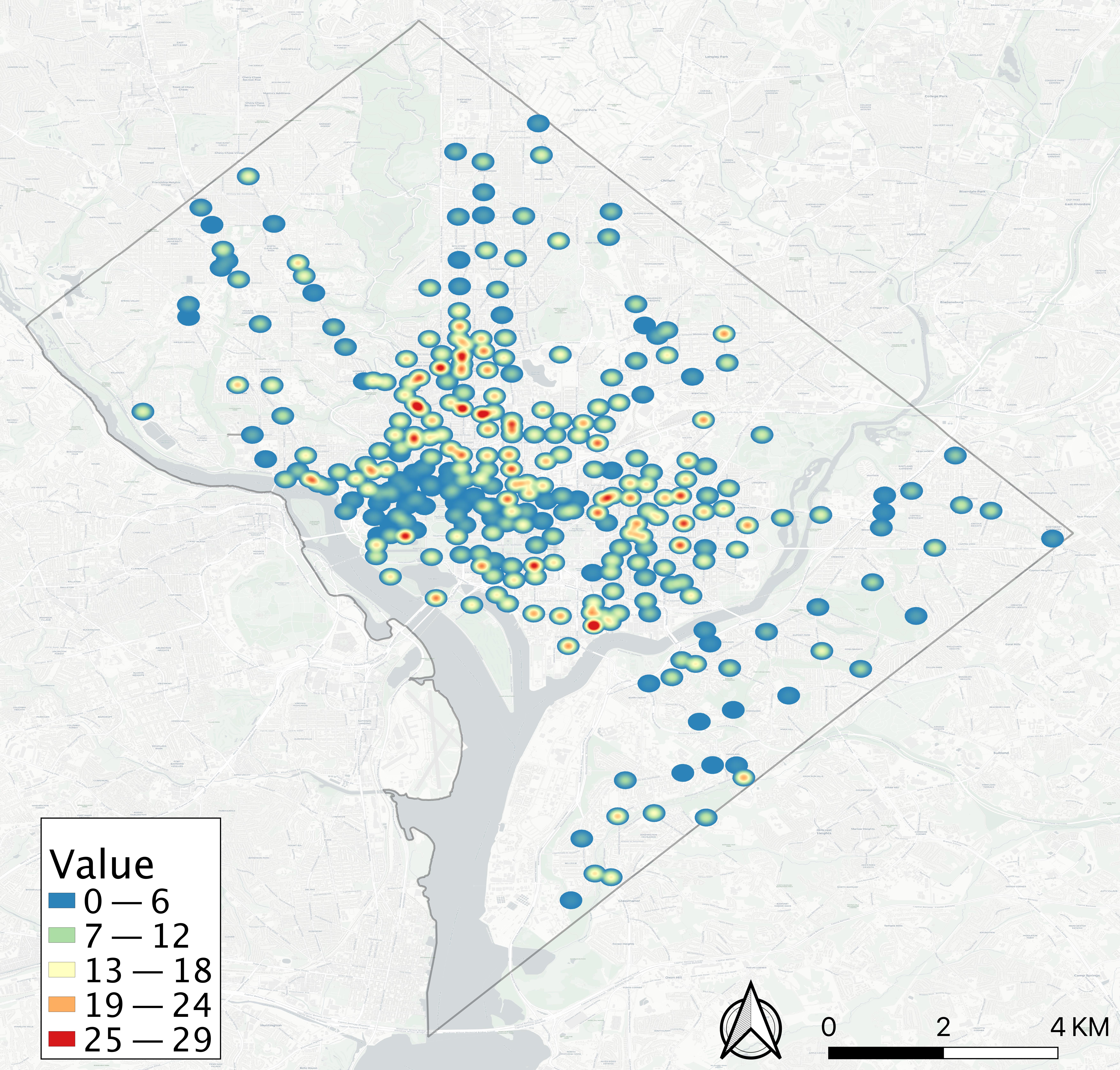}
         \caption{Bikesharing, before COVID}
         \label{fig:1b}
     \end{subfigure}
     \hfill
     \begin{subfigure}[h]{0.32\textwidth}
         \centering
         \includegraphics[width=\textwidth]{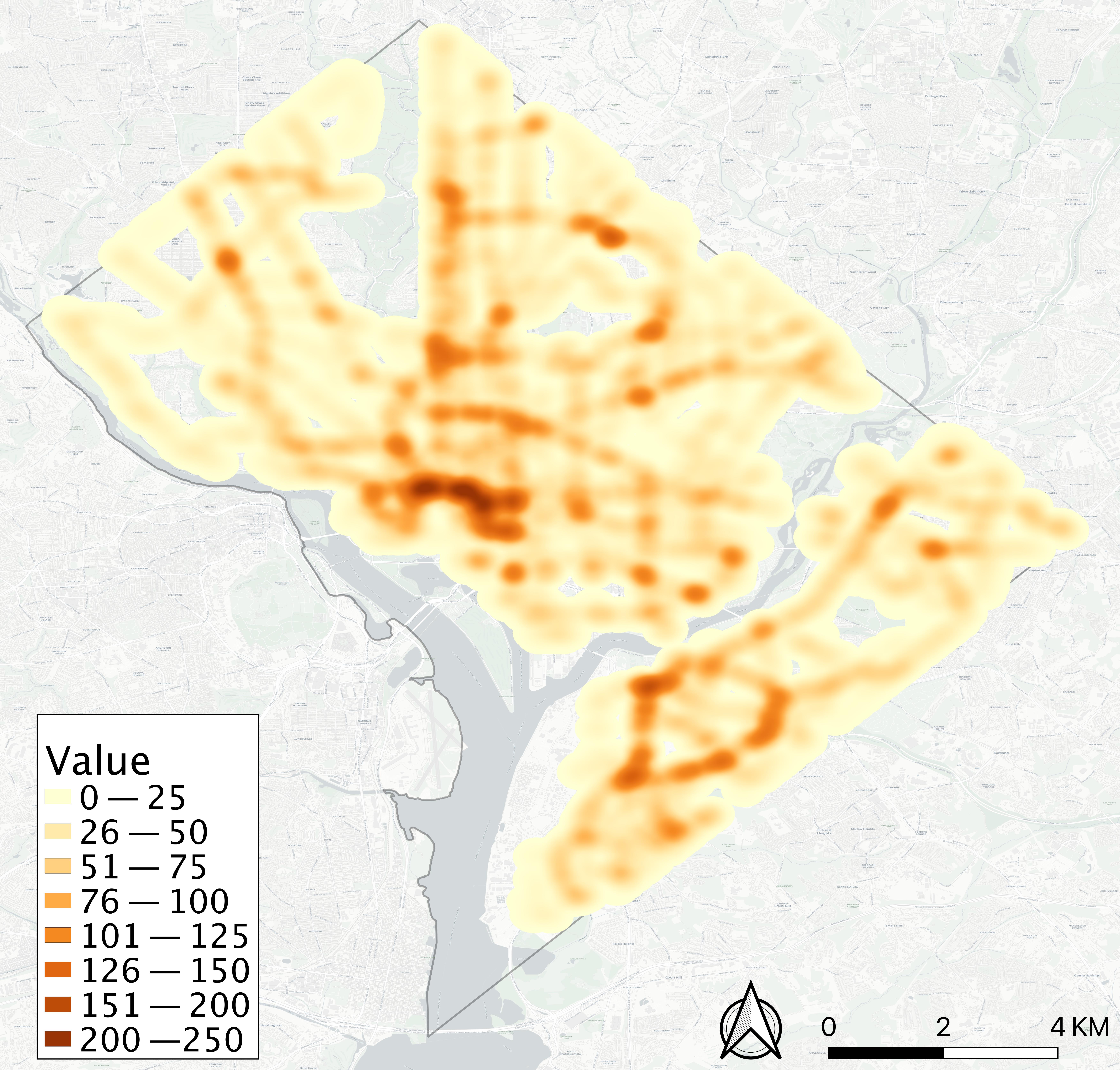}
         \caption{Transit, before COVID}
         \label{fig:1c}
     \end{subfigure}
      \hfill
          \begin{subfigure}[h]{0.32\textwidth}
         \centering
         \includegraphics[width=\textwidth]{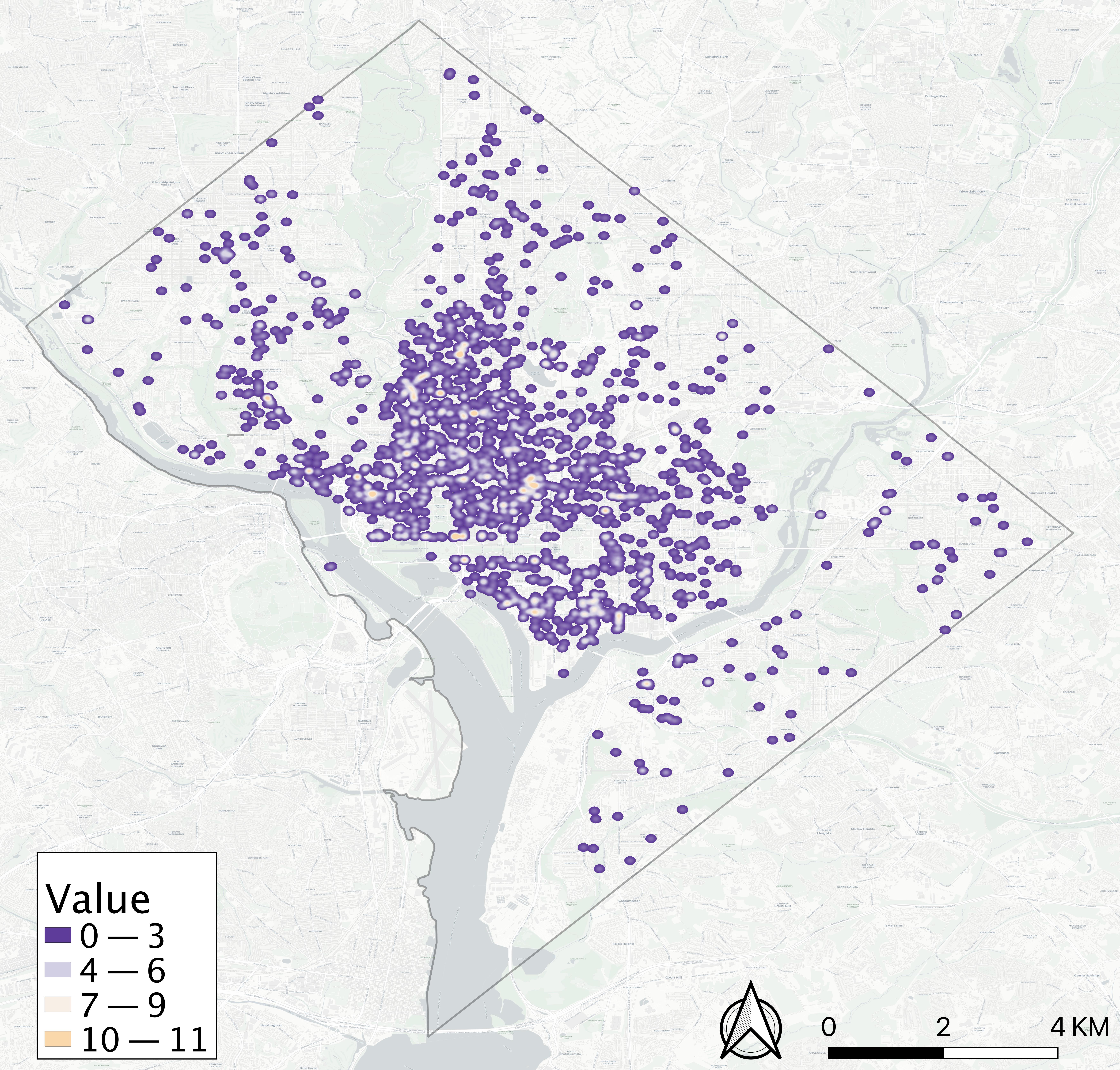}
         \caption{E-scooter, during COVID}
         \label{fig:2a}
     \end{subfigure}
     \hfill
          \begin{subfigure}[h]{0.32\textwidth}
         \centering
         \includegraphics[width=\textwidth]{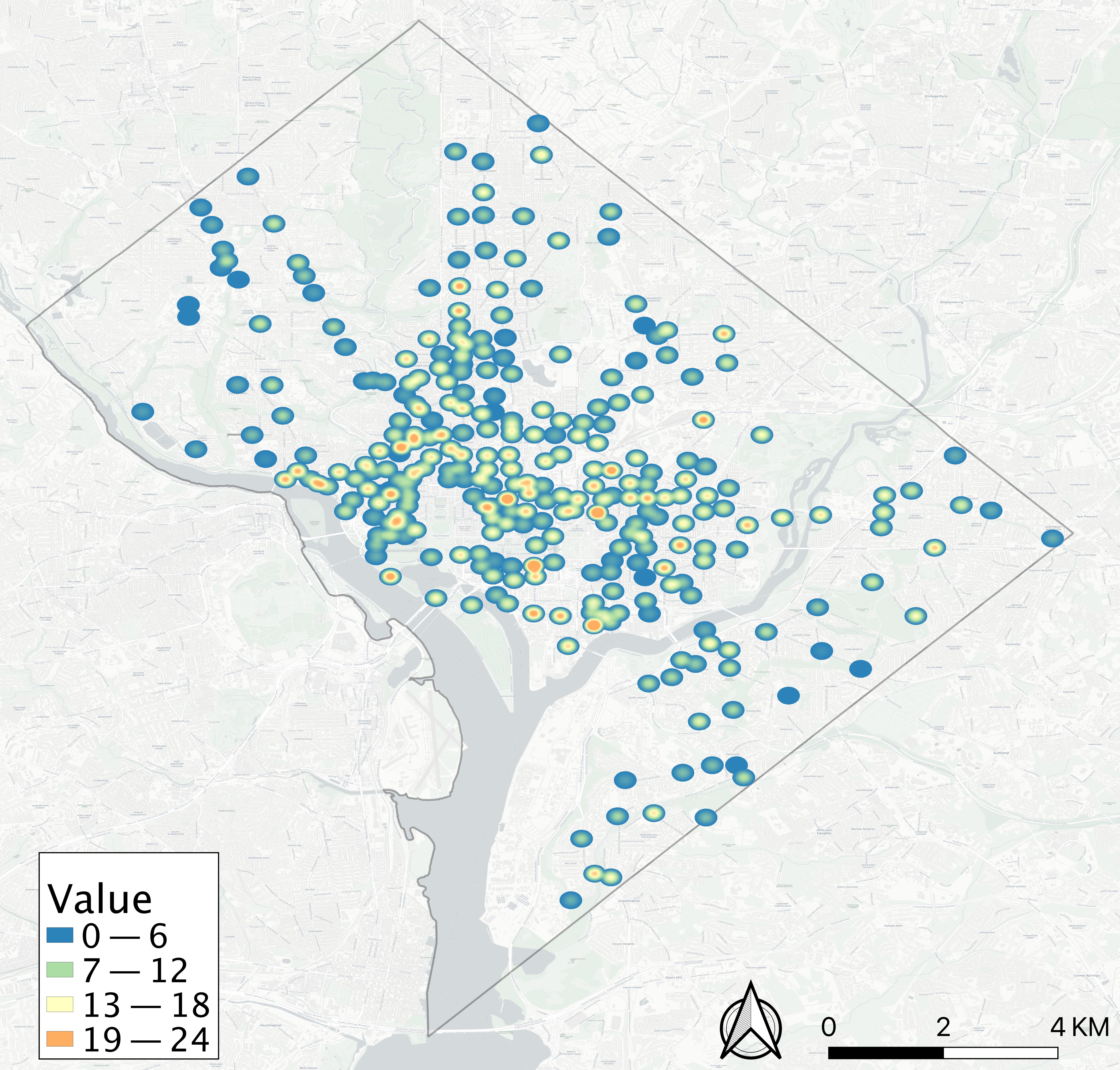}
         \caption{Bikesharing, during COVID}
         \label{fig:2b}
     \end{subfigure}
     \hfill
     \begin{subfigure}[h]{0.32\textwidth}
         \centering
         \includegraphics[width=\textwidth]{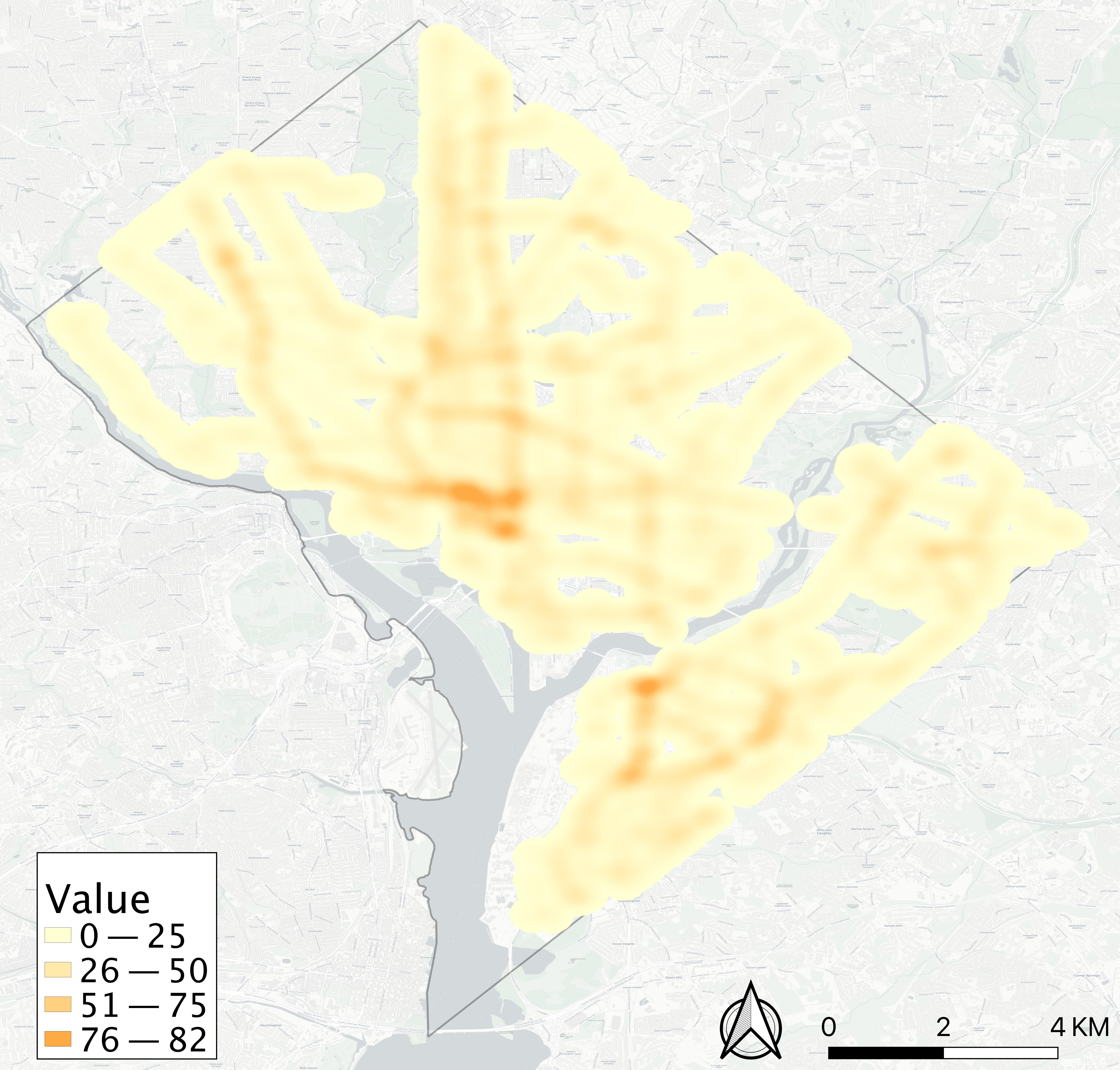}
         \caption{Transit, during COVID}
         \label{fig:2c}
     \end{subfigure}
     
        \caption{Kernel density estimation of e-scooter, bikesharing, and transit supply before and during COVID}
        \label{fig:1}
\end{figure}

Figure 2 shows the kernel density outputs for the before- and during-COVID week at 7:00 am, respectively. The spatial patterns are similar at other time points (i.e., 12:00 pm, 5:00 pm, and 8:00 pm), and so we do not present the results. Note that while the color ramp is the same for all maps, the corresponding value distributions are different. In other words, one should focus on the spatial patterns revealed by each map and not compare kernel density values across maps. These maps generate some useful insights. First of all, the spatial distribution of e-scooter supply is similar to that of bikesharing except two noticeable differences. One is that e-scooters are more spatially concentrated around the Downtown and Capitol areas (located at upper part of Ward 2), where more trips are likely to be generated. The other is that e-scooters are accessible to a wider geographic area than bikesharing. This illustrates a key advantage of e-scooters: their free-floating nature allow them to be deployed everywhere, providing a great potential to fill the services gaps left out by station-based systems. On the other hand, e-scooter services do not appear to expand the service area of public transit. This is because Washington DC has widespread transit coverage in the first place. Moreover, the supply intensity of transit services is more evenly distributed than that of e-scooters and bikesharing, which reflects the fact that the operation of transit services is less market-driven than the other two modes. Finally, the spatial patterns are largely similar for the pre-COVID week and the during-COVID week. Noticeably, the kernel density values for transit supply in the during-COVID week are much higher than those in the pre-COVID week, which indicate a significant service cut.


We further apply a statistical approach to complement the visual inspections. Specially, we use correlation analysis to evaluate the similarity of the spatial distributions of mobility-service supply intensity. We first develop a quarter-mile by quarter-mile fishnet for the city of Washington DC, based on which we then use zonal statistics to get the mean kernel density value at each cell. The obtained value indicates the supply intensity for a given mode at each cell. Finally, we compute the correlation coefficients between the supply-intensity values of the three modes at different hours (7:00 am, 12:00 pm, 5:00 pm, and 8:00 pm). A correlation coefficient closer to one (a value of one indicates a complete overlap of service provisions) indicates a stronger competing relationship, whereas a value closer to zero indicates greater complementary effects. 

The results are presented in Table \ref{tab:kerneldensitycorrelation}. At any point in time, the degree of competition is the strongest between bikesharing and e-scooters, whose correlation ranges from 0.60 to 0.73. Moderate level of competition exists between e-scooter services and public transit, as the correlation coefficients are in the range of 0.45 to 0.61. Finally, the correlation is the weakest between bikesharing and public transit, which indicates some level of coordination between the two systems. The District Department of Transportation has prioritized integrating bikeshare into the existing transit system as a major objective in development plans of Capital bikeshare.\footnote{See the recent District of Columbia Capital Bikeshare Development Plan at \url{https://ddot.dc.gov/sites/default/files/dc/sites/ddot/page_content/attachments/Draft\%20DDOT\%20Bikeshare\%20Development\%20FINAL\%20reduced.pdf}.} Before COVID-19, the competition between different travel modes appear to be more intense in the afternoon and in the evening. By examining the origins and destinations of e-scooter trips and destinations (results not presented), we find that this is largely due to imbalanced trip flows. That is, more e-scooter and bikeshare trips end at central locations (e.g., the downtown and Capitol areas) than they start from these locations, which shifts the supply of e-scooters and share bikes to be more concentrated. This finding indicates the need for rebalancing efforts throughout the day to ensure an equitable distribution of e-scooter and bikesharing services. By contrast, the trip flows between different zones of the city are quite balanced for the during-COVID week, and so the correlation coefficients do not vary much by time of day in this week. Finally, the correlation coefficients in the during-COVID week are generally smaller than those in the pre-COVID week, indicating more complementarity between modes during COVID-19 when transit services are significantly reduced.

\begin{table}[!ht]
\caption{Kernel Density Correlation Coefficients}\label{tab:kerneldensitycorrelation}
\centering
\small
\begin{tabular}{c|ccc|ccc}
\hline
\multicolumn{1}{l|}{} & \multicolumn{3}{c|}{Before COVID}                                                                                                                                                                         & \multicolumn{3}{c}{During COVID}                                                                                                                                                                         \\
Hour                 & \begin{tabular}[c]{@{}c@{}}Bikeshare \&\\  E-scooter\end{tabular} & \begin{tabular}[c]{@{}c@{}}Transit \&  \\  E-scooter\end{tabular} & \begin{tabular}[c]{@{}c@{}}Bikeshare \& \\  Transit\end{tabular} & \begin{tabular}[c]{@{}c@{}}Bikeshare \&\\  E-scooter\end{tabular} & \begin{tabular}[c]{@{}c@{}}Transit \&  \\  E-scooter\end{tabular} & \begin{tabular}[c]{@{}c@{}}Bikeshare \& \\  Transit\end{tabular} \\ \hline
7:00 AM              & 0.66                                                              & 0.48                                                              & 0.37                                                             & 0.62                                                              & 0.45                                                              & 0.38                                                             \\
12:00 PM             & 0.72                                                              & 0.58                                                              & 0.50                                                             & 0.60                                                              & 0.46                                                              & 0.41                                                             \\
5:00 PM              & 0.73                                                              & 0.61                                                              & 0.53                                                             & 0.61                                                              & 0.47                                                              & 0.40                                                             \\
8:00 PM              & 0.69                                                              & 0.54                                                              & 0.46                                                             & 0.63                                                              & 0.45                                                              & 0.41                                                             \\ \hline
\end{tabular}
\end{table}



\subsection{Classification of E-Scooter Trips: Substitute or Complement}
To address \textbf{RQ2}, we conduct a spatiotemporal analysis of e-scooter trip origins and destinations, focusing on their relationships with transit and bikesharing stations. The analysis has two separate parts. First, we classify e-scooter trips into different types by examining whether their origins and destinations fall into the service area of transit or bikesharing stations. Second, we identify likely combined e-scooter and transit trips by examining if an e-scooter trip starts from or ends at a transit stop. 

\subsubsection{Classifying E-Scooter Trips} 
In this paper, we develop a new typology of e-scooter trips regarding their level of competition/complementarity with transit, as shown in Figure 3. The key distinguishing factor is whether the e-scooter trip ends locate within or outside the transit coverage area. Here, we define the service radius of a transit stop as a quarter mile (a 5-minute walk), a commonly used threshold by transit analysts \citep{walker2012human}. We use ArcGIS's Network Analyst to generate the service area (a 5-minute walking shed) for each transit stop. For the fist two trip types, both the e-scooter trip origins and destinations fall into the transit service area; however, the two transit stops are connected with a direct transit line for Trip Type 1 and they are not for Trip Type 2. In other words, the transit alternative to an e-scooter trip classified as type 2 involves at least one transfer. Hence, while both Trip Type 1 and 2 can be considered as competing with transit, the degree of competition is stronger for Trip Type 1. For Trip Type 3, one trip end locates within the transit service area and the other locates outside of it. E-scooter trips in this category are likely to either serve as a last-mile feeder to transit or fill a spatial gap of transit coverage. Finally, Trip Type 4 clearly fills a spatial gap in the transit network as both trip ends locate outside of the existing service area.

\begin{figure}[!ht]
\centering
\includegraphics[width=0.9\textwidth]{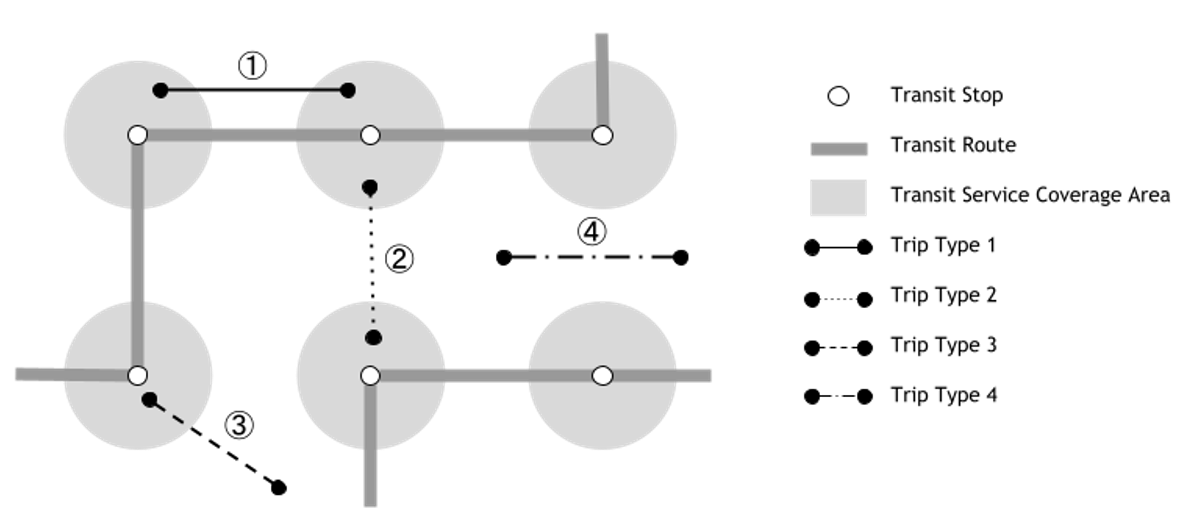}
\caption{Trip Classification}
\label{fig:tripclassification}
\end{figure}

Likewise, e-scooter trips can be classified into several types to indicate their degree of competition/complementarity with bikesharing. The classification differs from that with respect to e-scooter's relationship with transit in two key aspects. First, we set the service radius of a bikesharing station as one eighth of a mile, half of that of transit stops. Second, only three trip types are identified (the first two trip types should be merged), as bikesharing does not involve transfers. To distinguish the results from those concerning e-scooter's relationship with transit, We name them Trip Class 1, Trip Class 2, and Trip Class 3. 

Figure 4 presents the classification results regarding e-scooter's relationship with transit. Over 90\% of e-scooter trips are classified as Trip Type 1 or Trip Type 2. About 5\% of e-scooter trips have one of their trip ends fall outside of the transit service area, and very few trips (close to a zero percent of all trips) have both ends fall outside of the transit service area. These results are not surprising, as the transit network has covered the vast majority of the city's territory. Before one can readily interpret these results as suggesting a strong competitive relationship between e-scooters and transit, however, two additional considerations should be taken into account. First, e-scooter users are often not the same group of people who use transit. As mentioned above, scooter users tend to be younger, college educated, and have a medium to high income \citep{Alington2019,Portland2018,SF2019}. By contrast, transit riders come from all age groups and they tend to be poorer \citep{TransitRider2017}. Second, a significant proportion (about 20\%) of the trips are classified as Trip Type 2, which means that one cannot use transit to complete the same trip without taking a transfer. If the use of an e-scooter results in significant travel-time savings compared to the transit alternative, simply classifying these trips as substituting transit is inappropriate. 

\begin{figure}[!ht]
     \centering
     \begin{subfigure}[h]{0.48\textwidth}
         \centering
         \includegraphics[width=\textwidth]{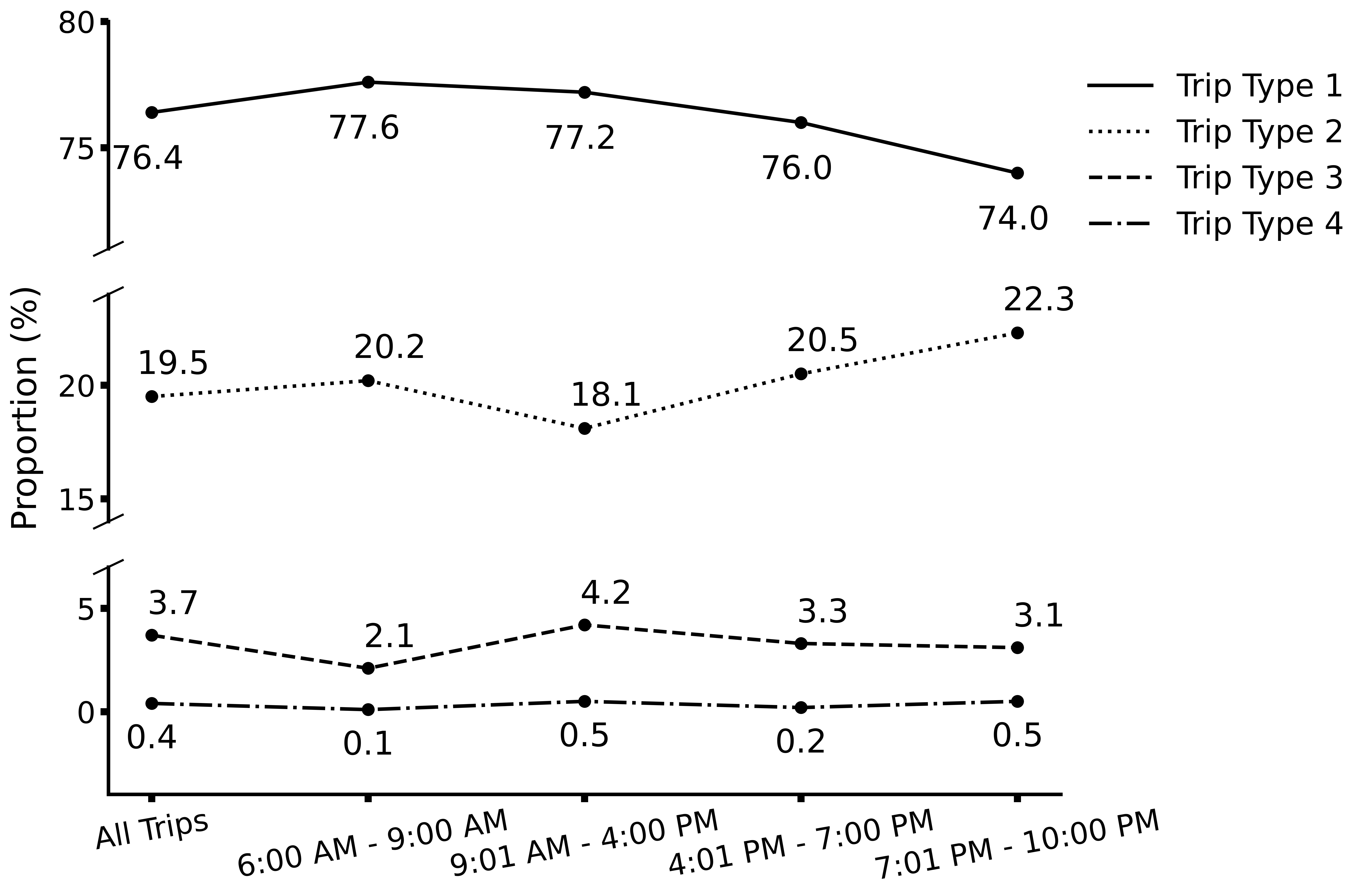}
         \caption{Relationship with transit, before COVID}
         \label{fig: transit before}
     \end{subfigure}
     \hfill
          \begin{subfigure}[h]{0.48\textwidth}
         \centering
         \includegraphics[width=\textwidth]{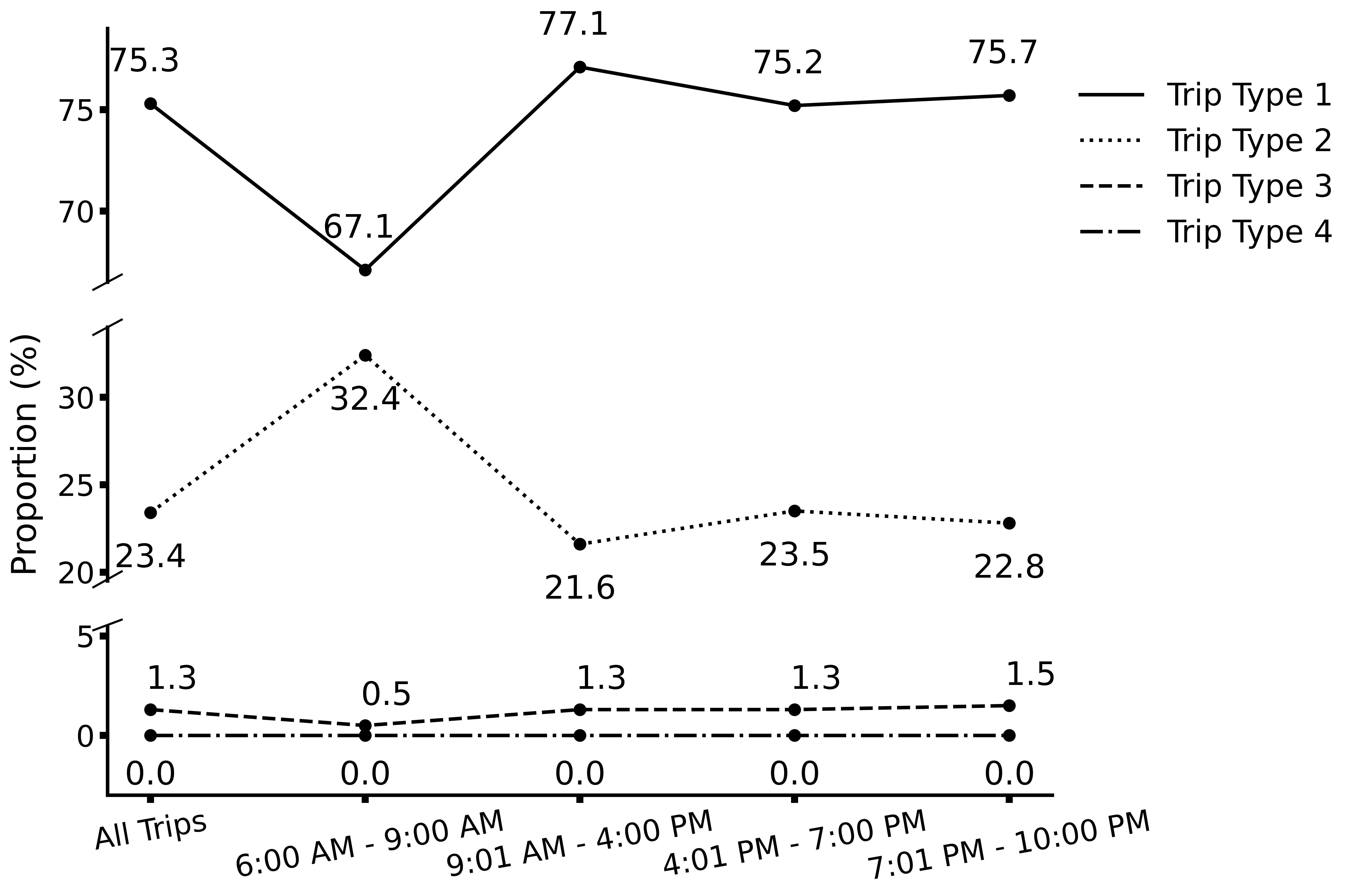}
         \caption{Relationship with transit, during COVID}
         \label{fig: transit during}
     \end{subfigure}
     
      \begin{subfigure}[h]{0.48\textwidth}
         \centering
         \includegraphics[width=\textwidth]{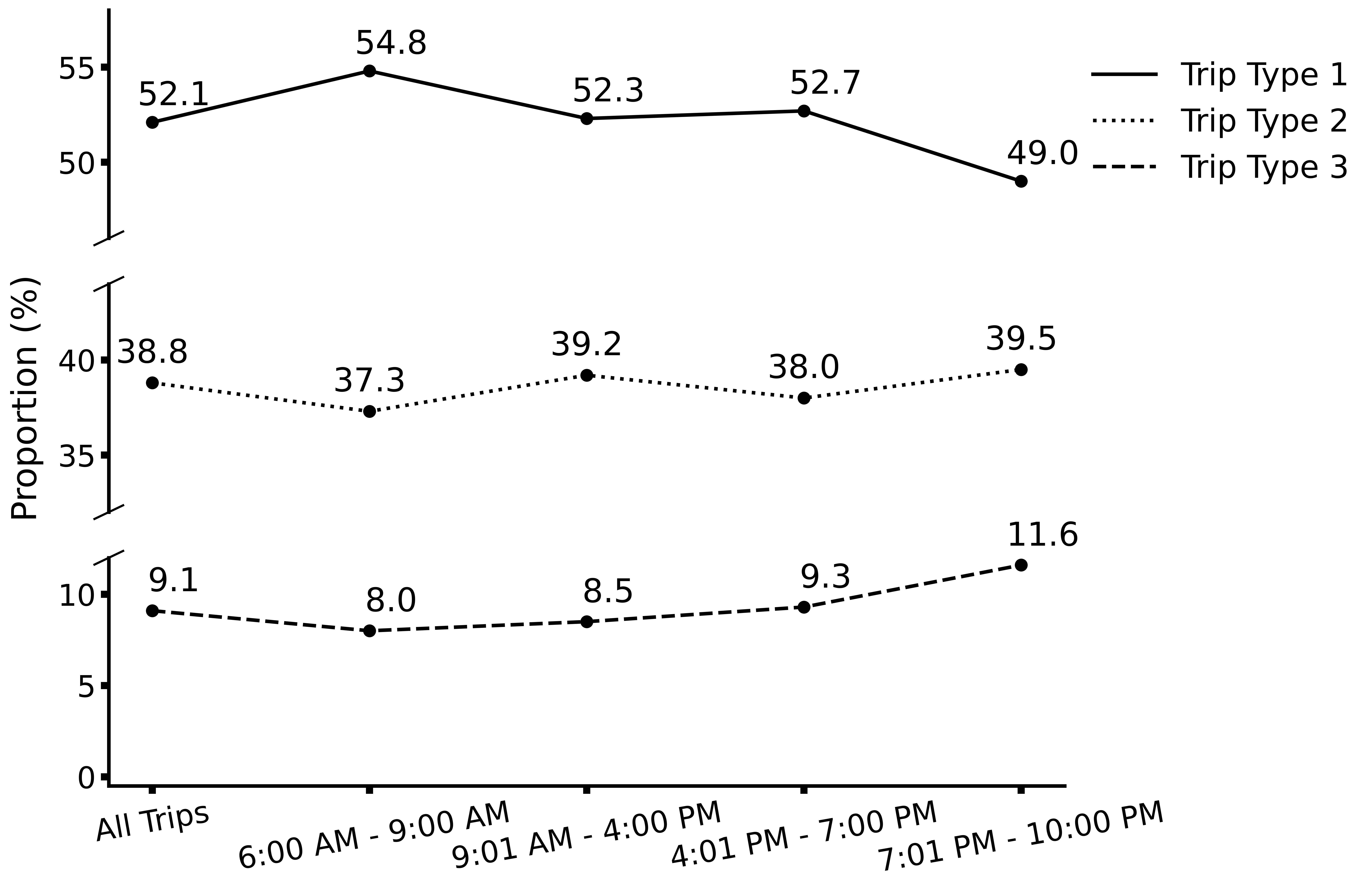}
         \caption{Relationship with bikesharing, before COVID}
         \label{fig: bikes before}
     \end{subfigure}
     \hfill
          \begin{subfigure}[h]{0.48\textwidth}
         \centering
         \includegraphics[width=\textwidth]{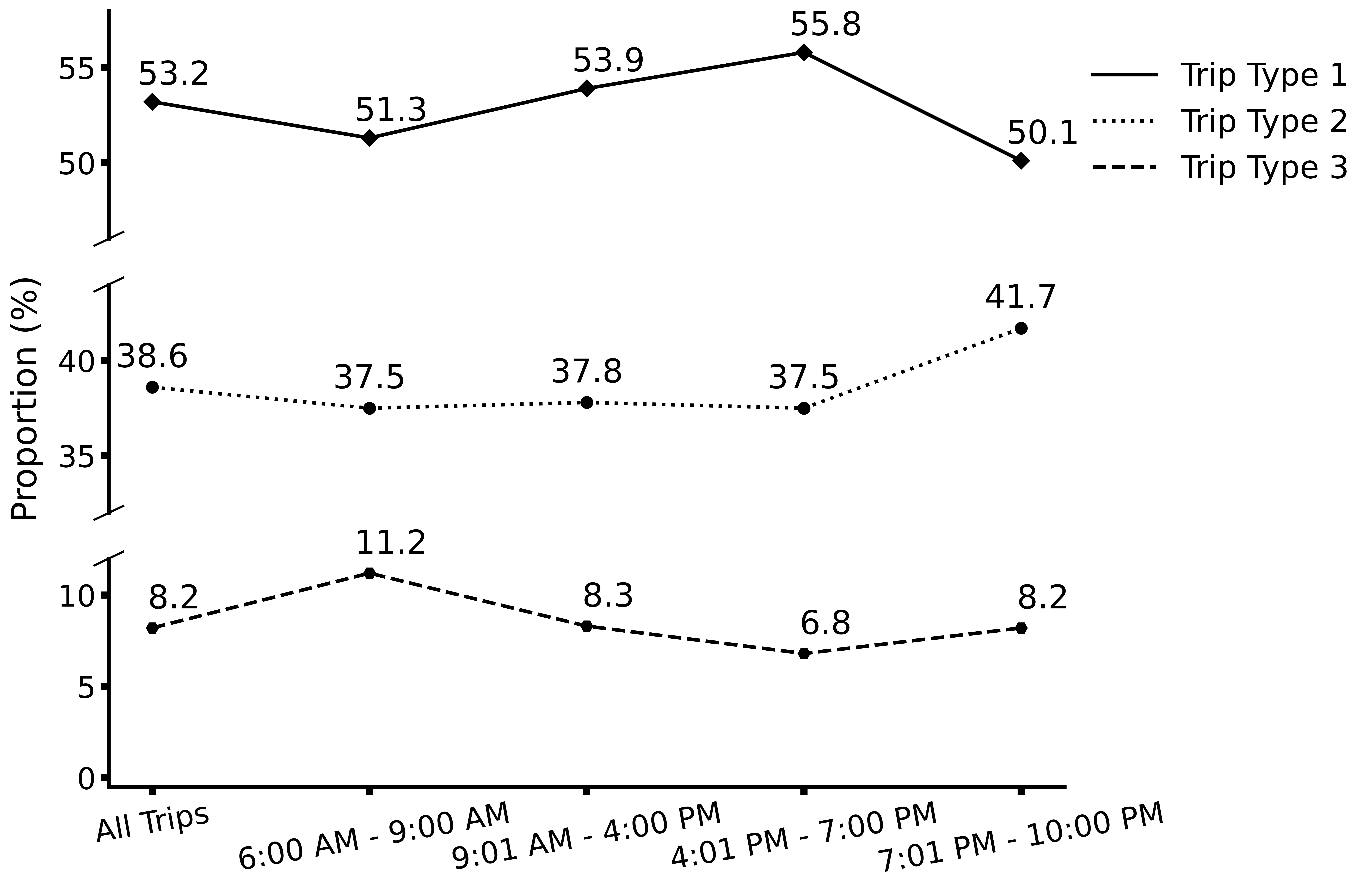}
         \caption{Relationship with bikesharing, during COVID}
         \label{fig: bikes during}
     \end{subfigure}
     
        \caption{E-scooter Trip Classification Results}
        \label{fig: classification results 1}
\end{figure}

The distribution of e-scooter trips across the four trip types differs slightly throughout the day, and the patterns are different for the before- and during-COVID week. Before COVID-19, a slightly higher proportion of trips are classified as Trip Type 1 or Trip Type 2 during the morning peak hours (6:00 am to 9:00 am), which is likely because many people use e-scooters for commuting to transit-rich destinations. By contrast, during COVID-19, a much lower proportion of e-scooter trips happening in the morning peak are classified as Trip Type 1. A plausible reason is the significant reduction of activities in central areas served by direct transit lines and especially the decreases in commuting trips to the downtown. Another significant difference between the before- and during-COVID periods is the decrease in the proportion of e-scooter trips being classified as Trip Type 3 or Trip Type 4. While this finding suggests a strong substitution effect of e-scooters on public transit, it also indicates that e-scooters may have played a important role in supporting transportation during a pandemic crisis when virus-wary travelers stay away from public transit.

Regarding e-scooter's relationship with bikesharing, we observe minor differences in the trip type distribution throughout the day and between the pre-COVID week and the during-COVID week. Most e-scooter trips (between 55 to 60 percent) happen at locations where bikesharing services are available, suggesting a strong competitive relationship. Previous research further suggests that e-scooters mainly compete with causal bikeshare use, as the spatiotemporal patterns of bikeshare trips taken by members are remarkably different from those of e-scooter trips \citep{mckenzie2019spatiotemporal}.

\subsubsection{Identifying Transit-connecting E-scooter Trips}
A special case for e-scooters to complement public transit is when they serve as a last-mile connection to transit. If an e-scooter trip starts or ends at a location next to a transit stop, it is likely a leg of a combined scooter-and-transit trip. Since previous studies on bikesharing find that travelers often use shared bikes to connect with rail services but not bus services, we focus on rail entrances only \citep{martens2004bicycle,martin2014evaluating}. We assume e-scooter trips that happen within a distance threshold of a rail entrance as potential integrated e-scooter-and-rail trips. Regardless of the distance threshold chosen here, some e-scooter trips will be falsely labelled and the bias can go both directions. A upward bias happens when trips falling within the distance threshold are not a leg of an assumed ``e-scooter plus transit'' trip, and a downward bias happens when transit riders park the e-scooter at a distance beyond the chosen threshold. Given these uncertainties, we use 30 feet as the threshold to get a lower bound estimate and use 100 feet to get a higher bound estimate.


In the pre-COVID week, we estimate that between 1174 and 1489 e-scooter trips are potentially connecting to metro, 8\% to 12\% of all trips. As people stay away public transit during COVID-19, however, both the number and proportion of transit-connecting e-scooter trips declined significantly. In this week, the estimated number of transit-connecting trips is between 6\% to 7\% of all trips. We further present the number of estimated transit-connecting trips by time of day in Figure 5. The graph shows that more combined scooter-and-transit trips happen during the peak hours, which indicate the use of e-scooters to facilitate commute trips by transit. Notably, in the pre-COVID week, about 20\% of e-scooter trips made in the morning peak hour are identified as transit-connecting trips. Therefore, as more and more cities embrace shared e-scooters services, commuting trips should be the main focus for transportation officials to promote e-scooters as a last-mile enhancement to public transit. 


\begin{figure}[!ht]
     \centering
         \includegraphics[width=\textwidth]{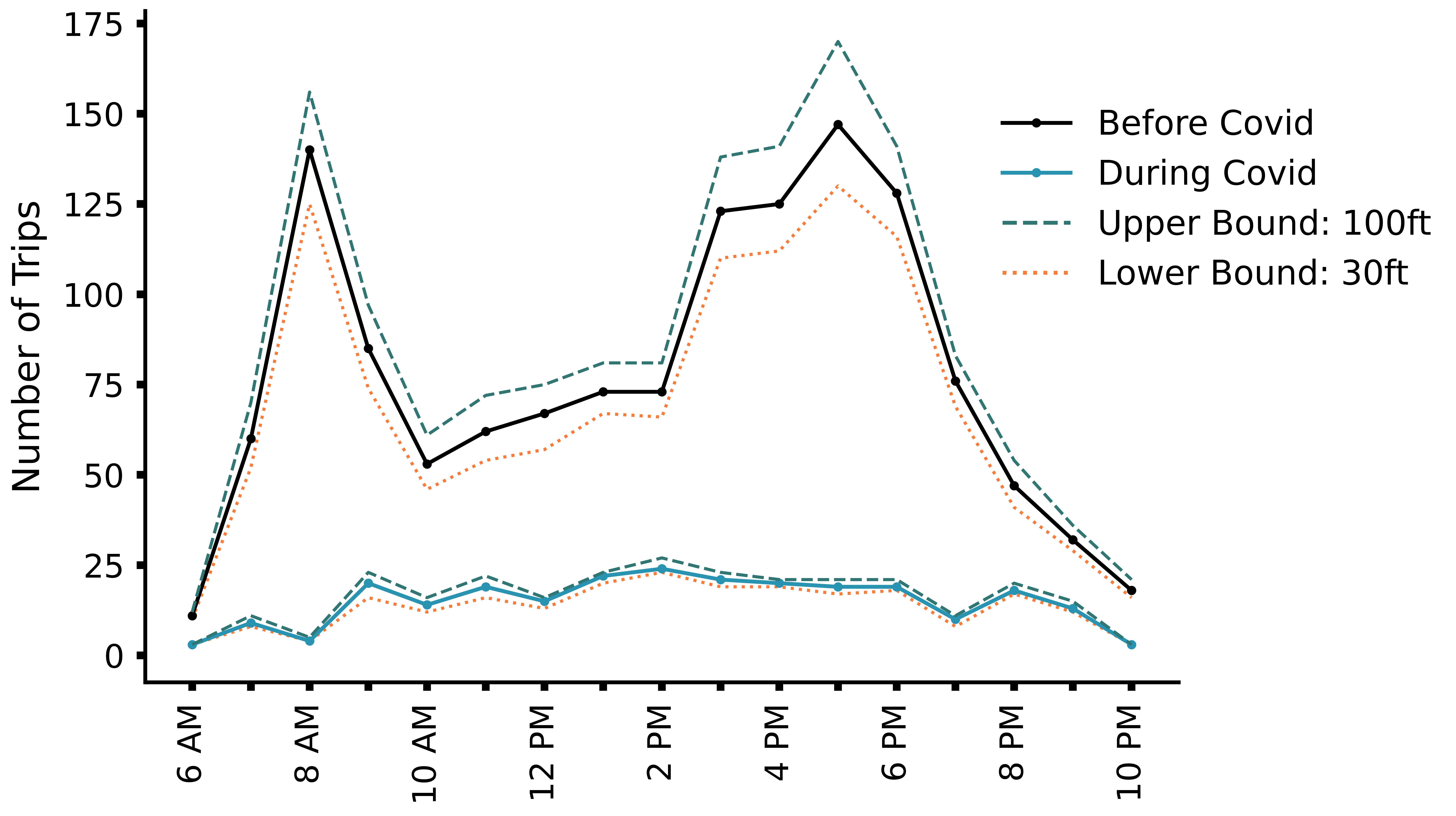}
     
        \caption{Number of transit-connecting trips identified before and during COVID}
        \label{fig: last mile updated}
\end{figure}

\subsection{Travel-Time Differences Between E-Scooter Trips and The Transit Alternative}
In this section, we address \textbf{RQ3} by examining the travel-time differences between e-scooter trips that are classified as Trip Type 1 and Trip Type 2 and their fastest transit alternative, which can shed light on \textit{why} people have choose to use e-scooters over transit. In this analysis, We exclude likely leisure e-scooter trips (i.e., taking e-scooters just for fun), which include the following: trips with an estimated travel speed lower than eight miles per hour, trips with a distance under a quarter mile, and trips happening at tourist sites such as the National Mall area. 

We generate the fastest transit alternative for each e-scooter trip using the ArcGIS Pro Network Analyst tool. The network is built based on the GTFS data. The before- and during-COVID weeks have different service schedules and hence different GTFS versions. We thus construct two separate networks for each week. Transit travel time includes the following components: walking time to and from transit stops, waiting time, boarding/alighting time (set as half of a min), in-vehicle travel time, and any time required for transfers if necessarily. The ArcGIS Pro Network Analyst tool outputs the total transit travel time by accounting for all of these time components. It does so for a user-specified date and departure time based on the transit schedules specified in the GTFS data. For each e-scooter trip, the departure time of its transit alternative should be approximate to the start time of the e-scooter trip. Here, we estimate and get the median of the transit travel times for every minute of the 10 minutes before and after the start time of the e-scooter trip. This accounts for the possibility that individuals may slightly adjust their schedule to minimize wait time, that is, they do not depart for transit exactly at the time they begin the e-scooter trip.

\begin{table}[]
\caption{Characteristics of e-scooter trips and their fastest transit alternatives}\label{tab:kerneldensitycorrelation}
\resizebox{1\textwidth}{!}{
\begin{tabular}{llll}
\hline
                                                       & Before-COVID & During-COVID & Significance \\ \hline
Median e-scooter trip length                           & 0.74 mile    & 0.95 mile    & ***          \\
Median e-scooter trip duration                         & 10 min       & 14 min       & ***          \\
Estimated travel time for transit alternative & 14.7 min     & 16.5 min     & ***          \\
\qquad (Travel time difference)                                 & (4.7 min)      & (2.5 min)      & ***          \\
Estimated median e-scooter trip cost                   & \$3.70       & \$6.00       &              \\
\% Trips made in morning peak (6 to 9 am)              & 8.6\%        & 4.8\%        &              \\ 
Number of e-scooter trips                                        & 12886        & 3825         &    \\ \hline
\end{tabular}
}
\begin{flushleft}\small{Note: Significance indicates if the difference between the before- and during-COVID weeks are statistically significant. *p<0.1, **p<0.05, ***p<0.01.} 
\end{flushleft}
\label{lastmile}
\end{table}

Table \ref{lastmile} presents the characteristics of e-scooter trips and their fastest transit alternatives in the before- and during-COVID weeks. As discussed above, e-scooters happening in the COVID-19 period are generally longer in distance and duration. While the median trip length is 0.74 miles and the median trip duration is 10 min for the before-COVID week, the median trip length is 0.95 miles and the median trip duration is 14 min for the during-COVID week. We find that the estimated travel times by transit are in generally greater than the e-scooter trip time for both periods. On average, an e-scooter trip's fastest transit alternative would take about 4.7 min longer than the e-scooter trip before COVID, but this travel-time difference decreases to 2.5 min during COVID even though transit service frequency is lower in this period. The decrease in travel-time difference between an e-scooter trip and its fastest transit alternative during the COVID period is likely attributable to reduced congestion levels.

A comparison of the estimated travel-cost differences between e-scooters and public transit sheds further light on travelers' choice behavior. The estimated median cost of e-scooter trips happening in the before-COVID week and the during-COVID week is \$3.7 and \$6.0, respectively.\footnote{A recent Washington Post article has reported the price information of e-scooter services in Washington DC \citep{DC2019Price}. The per-minute charge of e-scooter use has increased from 15 cents per minute in July 2019 to about 25 cents per minute in June 2020.} By contrast, the cost of a transit alternative is either \$2 (bus) or \$2.25 (rail) for both time periods. These numbers suggest that before COVID, e-scooter users often pay a small price premium (about \$1.5) to save a few minutes of travel time. During COVID-19 when e-scooters become more expensive to use, however, e-scooter users pay a much higher price premium (about \$4) for choosing e-scooters over transit but they saved very little time. These results indicate a significant role played by the COVID-19 virus: e-scooters have a stronger substitution effect on transit use during COVID when virus-wary travelers stay away from taking public transit. A further implication is that as transit agencies seek to recover from the COVID-19 pandemic crisis, focusing on easing traveler's public health concerns can be more effective than offering fare discounts.

\section{Discussion}
Overall, the results presented above indicate that e-scooters both compete and complement bikesharing and public transit. Evidence from both the supply side (service availability) and the demand side (mode use) supports this statement. On the supply side, we find that e-scooters compete with bikesharing at many locations, but they also extend micromobility access to neighborhoods where bikesharing is unavailable. Since Washington DC has widespread transit coverage, e-scooters do not appear to extend the service areas of public transit. Nevertheless, during COVID-19 when transit services are significantly reduced, the supply of e-scooters has a more complementary spatial relationship with public transit.

The trip classification results (demand-side analysis) are consistent with the findings from the supply-side analysis. We find that a majority of e-scooter trips can be classified as substitutes to bikesharing and public transit. Before COVID-19, some travelers pay a slightly higher price to use e-scooters in order to save some travel time as opposed to riding transit. However, we also find some evidence of a complementary effect, as travelers use e-scooters to reach locations where bikesharing and transit stations are inaccessible by walking. Notably, e-scooters appear to serve some essential trips, albeit at a higher price, during COVID-19 as virus-wary travelers stay away from buses and trains. In addition, a sizeable proportion of e-scooter trips are taken to connect with the transit system, which indicates the potential for e-scooters to address the last-mile problem of public transit. 


These findings have major implications for transportation planning and policymaking. First of all, as e-scooters grow in popularity, cities are expected to see some disruptive impacts on the existing public transportation options. At present, the volume of e-scooter trips is relatively small compared to the volumes of bikeshare and transit trips (the latter are dozens and hundreds of times larger in scale according to our estimates), which explains why e-scooter trips was found not to affect Capital Bikeshare in a Phase 1 Evaluation study conducted in late 2018 \citep{DC2018Pilot}. However, if e-scooters services keep growing, they can attract away some bikeshare and transit customers who enjoy scooter rides and are more willing to pay. Results presented here show strong spatial competition effects between e-scooters and existing mobility options. Previous studies showed that younger adults are more likely to switch from station-based bikesharing to dockless bikesharing and that more substitution happens for short-duration trips \citep{li2019London,li2019Nanjing}. As city officials develop plans for bikesharing and transit systems, a careful market analysis should be conducted to assess the influences of e-scooters.

Moreover, results from the spatiotemporal analysis of service availability and use inform policy guidelines on e-scooter deployment and operations. To ensure accessible and equitable mobility services to all travelers, cities should incentivize e-scooter companies to place vehicles at neighborhoods unserved or underserved by existing mobility options. Some cities have already been doing so by requiring permit holders to place a certain number or proportion of vehicles in city-specified priority zones or equity zones at the start of the day \citep{NACTOGuide2019}. Our study suggests that this requirement can be strengthened in two aspects. First, besides the commonly considered equity criteria, the definition of the priority or equity zones should incorporate a consideration of the availability of existing mobility options. The kernel density approach adopted here is a plausible way to quantify the supply intensity of different mobility options. This approach allows cities to identify service gaps at a high spatial resolution. The size of the priority or equity zones should not be too large, otherwise e-scooter companies may cluster e-scooters at some locations while leaving other parts of the zone unserved. Second, rebalancing efforts are required to ensure equitable distribution of service availability throughout the day. We finds that e-scooters become less complementary to bikesharing and transit in later hours of day due to imbalanced trip flows.

In addition, e-scooter services can contribute to the resiliency of a public transportation system. COVID-19 has caused transit ridership to plummet as virus-wary travelers stay way public transit \citep{liu2020impacts}. As this effect lingers, one would hope that more sustainable travel modes rather than personal vehicles can replace the foregone transit trips. And if public agencies cut transit services due to a declining ridership, cities need to support affordable mobility options as an alternative. E-scooters, a personalized and relatively affordable (especially if a low-income program is offered) travel option, are demonstrated here a viable substitute to public transit during COVID-19. As cities reopen the economy and people resume social activities, e-scooters and other micromobility options can play a major role in helping people get around. This means the need for cities to develop appropriate bicycle and parking infrastructure to accommodate their use. 

\section{Conclusion}
This paper examines the spatiotemporal patterns of e-scooter service availability and use in Washington DC, focusing on analyzing if e-scooters complement or substitute public transit and bikesharing. The open data scrapped from publicly available APIs allow us to investigate this issue from both the supply and demand side. Existing research has rarely examined if and how much the supply of e-scooter services overlap with that of other modes. 


We find that e-scooters have both substituting and complementary effects on bikesharing and public transit. E-scooters have mostly been placed at locations where transit and bikesharing services are also available, which indicates that they often compete for the same customer base. Based on the spatial locations of e-scooter trip origins and destinations, we classify a majority of e-scooter trips as substitutes to transit and bikesharing. We further show that a majority of e-scooter trips take a shorter time than their fastest transit alternatives. On the other hand, e-scooters also complement existing mobility options. E-scooter expand access to neighborhoods without bikesharing services. Moreover, many travelers have used e-scooters for last-mile connection to public transit. 

We should note that the analysis conducted here is based on the Washington DC context, a city with one of the best public transit and bikesharing systems in the country. The maturity of these systems leaves little room for a new mobility option to supplement the existing services and to fill their service gaps. Therefore, we expect one to find greater complementary effects of e-scooter services on public transit and bikesharing in places with less robust public transport systems. In other words, we expect the findings of this study to be only generablizable to cities with robust transit and bikesharing systems. 


While the analysis of ``big data'' scrapped from public APIs generates valuable insights into how e-scooters fit into the existing transportation system, this approach has some major limitations. Notably, the e-scooter trip data are not associated with any demographic and socioeconomic information; that is, we have no knowledge of the people who have made these e-scooter trips. This limitation prevents us from examining how the use of e-scooter services differs across population groups. Moreover, the analysis of traveler's preferences and behaviors is largely absent in this study. Behavioral questions such as how people's use of different travel modes change after adopting e-scooters and under what conditions are people more likely to use e-scooters to connect with public transit are unexplored. The traditional ``small data'' approaches such as surveys, interviews, and focus groups are more suitable to address these questions. Future research should consider integrate big-data and small-data approaches to generate more accurate and more comprehensive knowledge on how e-scooter services interact with public transit and bikesharing.

\bibliographystyle{abbrvnat}
\biboptions{semicolon,round,sort,authoryear}
\bibliography{trb_template}
\end{document}